\documentstyle[bibnorm]{lamuphys}
\makeatletter
\let\chapter\hid@chapter
\makeatother
\newcommand{\bm}[1] {\vec{#1}}
\input epsfig.sty
\begin{document}
\pagenumbering{arabic}

\pagestyle{empty}
\Huge{\noindent{Istituto\\Nazionale\\Fisica\\Nucleare}}

\vspace{-3.9cm}

\Large{\rightline{Sezione di ROMA}}
\normalsize{}
\rightline{Piazzale Aldo  Moro, 2}
\rightline{I-00185 Roma, Italy}

\vspace{0.65cm}

\rightline{INFN-1236/98}
\rightline{November 1998}

\vspace{2.cm}

\begin{center}{\large{ \bf A Poincar\'e Covariant Current Operator for Deep Inelastic Structure Functions and Deuteron Electromagnetic Properties}}
\end{center}
\vskip 1em
\begin{center} F.M. Lev$^1$,  E. Pace$^2$ and G. Salm\`e$^3$ \end{center}

\noindent{$^1$\it Laboratory of Nuclear Problems, Joint Institute
for Nuclear Research, Dubna, Moscow region 141980, Russia}

\noindent{$^2$\it Dipartimento di Fisica, Universit\`a di Roma
"Tor Vergata", and Istituto Nazionale di Fisica Nucleare, Sezione
Tor Vergata, Via della Ricerca Scientifica 1, I-00133, Rome,
Italy} 

\noindent{$^3$\it Istituto Nazionale di Fisica Nucleare, Sezione
di  Roma I, P.le A. Moro 2, I-00185 Rome, Italy}

\vspace{1cm}

\titlerunning{A Poincar\'e Covariant Current Operator}

\begin{abstract}
In front-form dynamics the current operator can be constructed
from auxiliary operators, defined in a Breit frame where initial and final three-momenta of the system are directed along the $z$ axis. Poincar\'e covariance constraints reduce for auxiliary operators to the ones imposed only by kinematical rotations around the $z$ axis.
  Elastic and transition form factors can be extracted without any ambiguity and in the elastic case the continuity equation is automatically satisfied, once Poincar\'e, ${\cal P}$ and ${\cal T}$ covariance, together with  hermiticity, are imposed.
Applications to deep inelastic structure functions in an exactly solvable model and to the calculation of the deuteron electromagnetic form factors are presented. 
\end{abstract}
\vspace{2.5cm}
\hrule width5cm
\vspace{.2cm}
\noindent{\normalsize{Proceedings of the Workshop 
on "Electron Nucleus Scattering", Elba, June 22-26, 1998. }}

\newpage
\pagestyle{plain}

\section{Introduction}

Both  QCD theory and effective models use the current operator as a
fundamental input for evaluating elastic and inelastic form factors of
relativistic systems of interacting particles. The electromagnetic (em) and weak current operators for these systems should properly commute with the Poincar\'e generators and satisfy hermiticity. For instance, in the case of deep inelastic scattering (DIS), the hadronic tensor
\begin{equation}
W^{\mu\nu}=\frac{1}{4\pi}\int\nolimits e^{\imath qx} \langle P,\chi|
J^{\mu}(x)J^{\nu}(0) |P,\chi\rangle d^4x
\label{01}
\end{equation}
will have correct transformation properties relative to the Poincar\'e group
(i.e., $W^{\mu\nu}$ will be a true tensor) only if both the initial state of the nucleon, $|P,\chi\rangle$ with four-momentum $P$ and internal wave function $\chi$, and the operator $J^{\mu}(x)$ have correct transformation properties with respect to the {\em same} unitary representation of the Poincar\'e group.

The electromagnetic current should also satisfy parity, ${\cal P}$, and time reversal, ${\cal T}$, covariance, i.e., it has to satisfy extended Poincar\'e covariance, as well as continuity equation.
 
An example of the necessity of compatibility between the generators of the Poincar\'e group and the current operators is clearly met in the investigation of elastic and inelastic hadron form factors within the front-form Hamiltonian dynamics \cite{Dir}. In this framework hadron form factors are often calculated assuming  that, in the reference frame where $q^+=0$, the component $J^+(0)$ can be taken in impulse approximation, IA (the $\pm$ components of four-vectors are defined as $p^{\pm}=(p^0 \pm p^z)/\sqrt{2}$). Consider for example the elastic electron scattering for a spin $1$ system, as the deuteron or the $\rho$ meson. If $\lambda$ and $\lambda'$ are the helicities in the initial and final states, respectively, and $I_{\lambda'\lambda}=\langle \lambda'|J^+(0)|\lambda\rangle $, then, because of ${\cal P}$ and ${\cal T}$ covariance, all the matrix elements $I_{\lambda'\lambda}$ can be expressed in terms of $I_{11}$, $I_{00}$, $I_{10}$ and $I_{1,-1}$. As follows from extended Poincar\'e covariance, current conservation and
hermiticity, the elastic electron-deuteron scattering is described by three independent, real form factors and therefore the above matrix elements are not independent. As shown, e.g., in Refs. \cite{GrKo,CCKP}, if $\eta=Q^2/4m_d^2$, with $m_d$ the deuteron mass, then the following constraint, called "angular condition" must be fulfilled in the $q^+=0$
 frame, viz.
\begin{equation}
(1+2\eta)I_{11}-I_{00}-(8\eta)^{1/2}I_{10}+I_{1,-1}=0.
\label{1}
\end{equation}
However this relation, which is related to the rotational covariance of the current, is not satisfied if the matrix elements $I_{\lambda'\lambda}$ are calculated with the free operator, $J_{free}^+(0)$, and one has four independent matrix elements. Therefore different prescriptions are possible to calculate the three physical form factors and there is a large ambiguity in the theoretical results \cite{KS}.

 In Ref. \cite{LPS} it was shown that a particular attention has to be devoted to the choice of the reference frame. Indeed in the front form the rotations around the $x$ and $y$ axes are dynamical, while those
around the $z$ axis are kinematical. Because of this peculiarity, in the Breit frame where the momentum transfer, $\vec q$, is directed along the
spin-quantization axis (and $q^+ \not = 0$), all the requirements of extended Poincar\'e covariance can be satisfied by a current operator obtained from the free current.
 
  Using this Poincar\'e covariant current we investigated two distinct
problems: i) the DIS in an exactly solvable model, where the particles
interact through a relativistic harmonic oscillator potential; taking
exactly into account the interaction both in the initial and in the final state, the structure functions, averaged over small intervals of the Bjorken variable $x$, coincide in the Bjorken limit with the results of the parton model \cite{LPS1}; ii) the deuteron em properties; we will present in this paper our preliminary results for the deuteron elastic form factors.

\section{General Formalism}

Let $P$ be the operator of the four-momentum for the system under
consideration. As well-known (see, e.g., Ref. \cite{AB}), the current operator should satisfy the conditions
\begin{sloppypar}
\begin{equation}
\exp(\imath Px) J^{\mu}(0)\exp(-\imath Px)=J^{\mu}(x),
\label{31}
\end{equation}
\begin{equation}
U(l)^{-1}J^{\mu}(x)U(l)=L(l)^{\mu}_{\nu}J^{\nu}(L(l)^{-1}x)
\label{32}
\end{equation}

\noindent where $L(l)$ is the element of the Lorentz group corresponding to $l\in SL(2,C)$ and $U(l)$ is the representation operator corresponding to $l$.
\end{sloppypar}
Let us consider Eq. (\ref{31}) as the definition of $J^{\mu}(x)$ in terms of $J^{\mu}(0)$. Then the Poincar\'e covariance of $J^{\mu}(x)$ takes place if
\begin{equation}
U(l)^{-1}J^{\mu}(0)U(l)=L(l)^{\mu}_{\nu}J^{\nu}(0)
\label{33}
\end{equation}

In Ref. \cite{LPS} it was shown how obtain a current which satisfies
this Lorentz covariance condition. In Section 3 we will summarize our procedure, while in the present section a brief introduction of front form is given.
\begin{sloppypar}

Let $p$ be the particle 4-momentum, $g=p/m$ the particle 4-velocity, ${\bm s}$ the spin operator, and $\sigma$ the projection of the spin on the $z$ axis. Since $p^2=m^2$, only three components of $p$ are independent. In the front form one chooses ${\bm p}_{\bot}$ and $p^+$ as such components, where ${\bm p}_{\bot}\equiv (p_x,p_y)$ and  $p^{\pm}=(p^0\pm p^z)/\sqrt{2}$). 

For a general Lorentz transformation the states $|\phi \rangle$ transform according to 
 \begin{equation}
\langle p,\sigma|U(l)|\phi \rangle = 
\sum_{\sigma'} D_{\sigma\sigma'}^s[W(l,g')]\phi(p',\sigma')
\label{5}
\end{equation}
where $p'=L(l)^{-1}p$ and $W(l,g')$ is the front-form Wigner rotation
defined as

\begin{equation}
W(l,g')=\beta(g)^{-1}l\beta(g')
\label{Wigner}
\end{equation}

\noindent The matrices $\beta(g)\in$ SL(2,C) represent  the
front-form boosts. The action of the boost $L[\beta(g)]^{-1}$ is such that $p'=L[\beta(g)]^{-1}p \equiv({\bm p}'_{\bot}=0, p'^{+}=m/\sqrt{2},p'^{-}=m/\sqrt{2}$).
In Eq. (\ref{5}), $D^s(u)$ is the matrix of the unitary irreducible
representation (UIR) of the group $SU(2)$ with spin $s$, corresponding to $u\in SU(2)$ (it is easy to verify that $W(l,g')\in SU(2)$).

The space ${\cal H}$ for the representation of the Poincar\'e group describing a system of $N$ free particles with masses $m_i$ and spins $s_i$ $(i=1,2,...,N)$ can be realized as the space of functions
$\phi({\bm p}_{1\bot},p_1^+,\sigma_1,..., 
{\bm p}_{N\bot},p_N^+,\sigma_N)$.
Instead of the variables ${\bm p}_{1\bot}$, $p_1^+$,...,
${\bm p}_{N\bot}$, $p_N^+$, we consider the variables ${\bm P}_{\bot}$,
$P^+$, and the internal variables ${\bm k}_1$,..., ${\bm k}_N$, where
$P=p_1+...+p_N$ is the total four-momentum, and ${\bm k}_i$
is the spatial part of the four-vector
\begin{equation}
k_i=L[\beta(G)]^{-1}p_i,
\label{15}
\end{equation}
with  $G=P/M_0$ and $M_0=|P|\equiv |P^2|^{1/2}$.

 The wave function of the state $|P',\chi'\rangle \in {\cal H}$ with
four-momentum $P'$ and internal wave function $\chi' \in {\cal H}_{int}$ can be written in the form 
\begin{eqnarray}
&&\langle {\bm P}_{\bot},P^+;{\bm k}_1,\sigma_1,...,{\bm k}_N,\sigma_N
|P',\chi'\rangle =
2(2\pi)^3P^{'+}\delta^{(2)}({\bm P}_{\bot}-
{\bm P}_{\bot}')\cdot\nonumber\\
&&\delta(P^+-P^{'+})\chi'({\bm k}_1,\sigma_1,...,{\bm k}_N,\sigma_N )
\label{76}
\end{eqnarray}

 The spin operator ${\bm S}$ for the system as a whole acts only through the variables of the space ${\cal H}_{int}$ and is unitarily equivalent to the spin operator in the conventional form (see, e.g., Refs. \cite{BarHal,Ter,riv}):
\begin{equation}
{\bm S}= R^{-1}
({\bm {\cal L}}+{\bm s}_1+...+{\bm s}_N) R
\label{21}
\end{equation}
where ${\bm {\cal L}}$ is the total internal orbital angular momentum
operator, $R = \{\prod_{i=1}^{N} D^{s_i}[v(\frac{k_i}{m_i})]\}$ and $v(g) \in SU(2)$ is the so called Melosh matrix \cite{Mel}, which in the given context was first considered in Ref. \cite{Ter}.

The internal wave function can be written as follows 
\begin{equation}
\chi'( {\bm k}_1,\sigma_1,...,{\bm k}_N,\sigma_N ) =
\langle {\bm k}_1,\sigma_1,...,{\bm k}_N,\sigma_N |\chi'\rangle =
\langle {\bm k}_1,\sigma_1,...,{\bm k}_N,\sigma_N |R^{-1} |\Phi\rangle,
\label{12}
\end{equation}
where $|\Phi\rangle$ is an eigenstate of canonical spin, 
${\bm {\cal L}}+{\bm s}_1+...+{\bm s}_N$.
\end{sloppypar}
\section{Extended Lorentz covariance of the current operator and current
conservation} 
\label{S6}

\begin{sloppypar} 
Let $\Pi_i$ be the orthogonal projector onto the subspace
${\cal H}_i\equiv \Pi_i{\cal H}$ corresponding to the
eigenvalue of the mass operator equal to $M_i$ and to the eigenvalue
of the spin equal to $S_i$. In constituent quark models
the spectrum of the mass operator is discrete, but in the general case
one has also to consider the continuous spectrum (e.g., in the
parton model).
 For this reason we will not specify whether the index enumerating the
eigenstates of the mass operator is discrete or continuous. In the latter case a sum over $i$ should be understood as an integration.

The key property of our procedure \cite{LPS} is the following spectral
decomposition of the current operator, viz.

\begin{equation}
J^{\mu}(0)=\sum_{ij}\Pi_i J^{\mu}(0)\Pi_j
\label{39}
\end{equation}

\noindent Because of this decomposition the operator $J^{\mu}(0)$ is fully defined by the set of operators in internal space 
\begin{eqnarray}
J^{\mu}(P_i;P_j')\equiv\langle {\bm P}_{\bot},P^+|\Pi_iJ^{\mu}(0)\Pi_j|
{\bm P}_{\bot}',P^{'+}\rangle 
\label{42}
\end{eqnarray}
corresponding to definite values of the masses.  At any fixed
values of $({\bm P}_{\bot}',P^{'+})$ and $({\bm P}_{\bot},P^+)$,
these operators act from ${\cal H}_{j,int}$ to ${\cal H}_{i,int}$,
where ${\cal H}_{i,int}=\Pi_i {\cal H}_{int}$.

First of all, let us consider the covariance with respect to continuous
Lorentz transformations. 
\end{sloppypar}

In order to investigate in detail the constraints imposed on
$J^{\mu}(P_i;P_j')$ by Lorentz covariance, it is convenient to consider the current in a Breit frame. The Breit frame is defined as the
reference frame where the initial and final momenta are
\begin{equation}
K_i=B(H_{ij})^{-1}P_i, \qquad K_j'=B(H_{ij})^{-1}P_j'
\label{46}
\end{equation}
In Eq. (\ref{46}) $H_{ij} \equiv (P_i+P_j')/|P_i+P_j'|$ and $B(H_{ij})$  denotes the Lorentz transformation $L[\beta(H_{ij})]$. The four-vectors $K_i$ and $K_j'$ in Eq. (\ref{46}) are such that
\begin{equation}
K_i^2=M_i^2, \quad K_j^{'2}=M_j^2, \quad {\bm K}_i+{\bm K}_j'=0
\label{47}
\end{equation}

 The current operator $J^{\mu}(P_i,P_j')$ can be defined through a front-form boost from the current in the Breit frame
\begin{equation}
J^{\mu}(P_i,P_j')=B(H_{ij})^{\mu}_{\nu}j^{\nu}({\bm K}_{i};M_i,M_j)
\label{49}
\end{equation}
where we use $j^{\nu}({\bm K}_{i};M_i,M_j)$ to denote $J^{\mu}(K_i,K_j')$, i.e. the current in the Breit frame. The latter can be defined in terms of the current in the special Breit frame where the three-momentum of the system, ${\bm K_i}$, is directed along the $z$ axis:

\begin{eqnarray}
&&j^{\mu}({\bm K_i};M_i,M_j)=L[r({\bm K_i})]^{\mu}_{\nu}
D^{S_i}[v(\frac{K_i}{M_i})^{-1}r({\bm K_i})]\cdot\nonumber\\
&&j^{\nu}(K{\bm e}_z;M_i,M_j)D^{S_j}[r({\bm K_i})^{-1}
v(\frac{K_j'}{M_j})]
\label{57}
\end{eqnarray}
where $r({\bm K_i})\in SU(2)$ is such that
$L[r({\bm K_i})]K {\bm e}_z={\bm K_i}$. 

It has been shown \cite{LPS} that the operator $J^{\mu}(0)$ is Lorentz
covariant if the current in the special Breit frame is covariant with respect to rotations around the $z$ axis, $u_z$, viz.
\begin{equation}
j^{\mu}(K{\bm e}_z;M_i,M_j)=L(u_z)^{\mu}_{\nu}
\exp(-\imath \varphi S^z_i)j^{\nu}(K{\bm e}_z;M_i,M_j)\exp(\imath
\varphi S^z_j)
\label{56}
\end{equation}
where $\exp(-\imath \varphi S^z_{i(j)})=D^{S_{i(j)}}(u_z)$.

 In the front form the rotations around the $z$ axis are interaction free, and therefore the continuous Lorentz transformations constrain the current $j^{\mu}(K{\bm e}_z;M_i,M_j)$ for a non-interacting system in the same way as in the interacting case. The same property holds for the covariance with respect to a reflection of the $y$ axis, $P_y$, and with respect to the product of parity and time reversal, $\theta$, which leave the light cone $x^+=0$ invariant, and therefore are kinematical. 
 Since the full space reflection is the product of ${\cal P}_y$ and a dynamical rotation around the $y$ axis by $\pi$, while ${\cal T}=\theta {\cal P}$, the current operator satisfies ${\cal P}$ and ${\cal T}$ covariance if it satisfies Poincar\'e covariance and covariance with respect to $P_y$ and $\theta$.

In conclusion for an interacting system the extended Lorentz covariance is clearly satisfied by a current composed in our Breit frame by the sum of free, one-body currents, because the constraints are the same for a non-interacting and an interacting system, viz. 

\begin{equation}
j^{\mu}(K\bm{e}_z;M_i,M_j)=
\langle 0,P^{+}|\Pi_i J_{free}^{\mu}(0) \Pi_j|0,P^{'+}\rangle 
\label{free}
\end{equation}
where
\begin{equation}
J_{free}^{\mu}(0) = \sum_{i=1}^{N} j_{free,i}^{\mu}
\label{free'}
\end{equation}
with N the number of constituents in the system.

 The property of hermiticity
\begin{equation}  
j^{\mu}(-\bm{K};M_j,M_i)=j^{\mu}(\bm{K};M_i,M_j)^*
\label{91}
\end{equation}
where $^*$ means the Hermitian conjugation in ${\cal H}_{int}$, is satisfied for 
$|{\bm K}|\neq 0$ if

\begin{eqnarray}
&&j^{\mu}(K{\bm e}_z;M_i,M_j)^*= \nonumber \\
&&L[r_x(-\pi)]^{\mu}_{\nu} D^{S_j}[r_x(-\pi)]
j^{\nu}(K{\bm e}_z;M_j,M_i) D^{S_i}[r_x(-\pi)])^{-1},
\label{60}
\end{eqnarray}
where $r_x(-\pi)$ represents a rotation by $-\pi$ around the $x$ axis.
It is worth noting that Eq. (\ref{60}) represents a non trivial
constraint  when $M_i=M_j$ (i.e., for elastic scattering).

The continuity equation $\partial J^{\mu}(x)/\partial x^{\mu} =0$ in terms of $J^{\mu}(0)$ reads
\begin{equation}
[P_{\mu},J^{\mu}(0)]=0
\label{35}
\end{equation}

\noindent and will be satisfied if
\begin{equation}
(P_i-P_j')_{\mu}J^{\mu}(P_i,P_j')=0.
\label{61}
\end{equation}
In our particular Breit frame, if $K\neq 0$, Eq. (\ref{61}) becomes
\begin{equation}
j^-(K\bm{e}_z;M_i,M_j)=-{\left [ M^2_{i} /( 2 K^+_{i})  - M^2_{j}
 / (2 {K'}^+_{j}) \right ] \over (K_i^+ - {K'}_j^+)} j^+(K\bm{e}_z;M_i,M_j)
\label{67}
\end{equation}
and then only $j^+(K\bm{e}_z;M_i,M_j)$ and $j^-(K\bm{e}_z;M_i,M_j)$ are
constrained by the continuity equation, while
$\bm{j}_{\perp}(K\bm{e}_z;M_i,M_j)$ remains unconstrained.
In the elastic case ($M_i=M_j=m$;  $S_i=S_j=S$) Eq. (\ref{67}) reads 
\begin{equation}
j^-(K\bm{e}_z;m,m)=j^+(K\bm{e}_z;m,m)
\label{minus'}
\end{equation}
This condition implies that $j_z(K\bm{e}_z;m,m)=0$. As shown in ref. \cite{LPS}, it is important to notice that, in the elastic case, the extended Lorentz covariance and hermiticity imply Eq. (\ref{minus'}), i.e., impose current conservation.

\section{Deep inelastic scattering in an exactly solvable model}
\label{S9}

 Let us consider the hadronic tensor for a system of internal state
$|\chi_0\rangle$, mass $m$ and initial momentum $P$ in the Breit reference frame,
with ${\bm P}_{\bot}={\bm q}_{\bot}=0$, $P_{z}=-P'_{z}=K>0$,

\begin{eqnarray}
&& W^{\mu\nu}=\frac{1}{4\pi}\sum (2\pi)^4\delta^{(4)} (P+q-P')
\langle \chi_0|j^{\mu}(K\bm{e}_z;m,M') |\chi'\rangle \cdot\nonumber\\
&& \langle \chi'|j^{\nu}(-K\bm{e}_z;M',m) |\chi_0\rangle
\label{86}
\end{eqnarray}
where the sum is taken over all possible final states with
four-momentum $P'$, internal wave function $\chi'$, and mass $M'$ \cite{LPS1}.

We consider a system of two different, interacting particles with the same mass, $m_0$, and spin $1/2$, in the ground state with internal wave function 
\begin{equation}
\chi_0({\bm k}_{\bot},\xi,\sigma_1,\sigma_2) =
\langle {\bm k},\sigma_1,\sigma_2|\chi_{0}\rangle =
\langle {\bm k},\sigma_1,\sigma_2|R^{-1} |\Psi_0\rangle
\omega({\bm k})^{1/2}, 
\label{5'}
\end{equation}
where
$\xi=p_1^+/P^+$, ${\bm k}_{\bot}={\bm p}_{1\bot}- 
\xi{\bm P}_{\bot}$, and $|\Psi_0\rangle$ is an eigenstate of canonical spin (see Eq. (\ref{12} and Ref. \cite{LPS1}). The internal momentum is ${\bm k}=({\bm k}_{\bot},k_z)$, where $k_z=(2 \xi - 1) \omega({\bm k})$ and $\omega({\bm k})=(m_0^2 + {\bm k}^2)^{1/2}$.

Because of hermiticity, Eq. (\ref{91}), it is possible to rewrite Eq. (\ref{86}) in the form 
\begin{eqnarray}
&& W^{\mu\nu}=\frac{1}{4\pi}\sum (2\pi)^4\delta^{(4)} (P+q-P')
\langle \chi_0|j^{\mu}(K\bm{e}_z;m,M') |\chi'\rangle \cdot\nonumber\\
&& \overline{\langle \chi_0|j^{\nu}(K\bm{e}_z;m,M') |\chi'\rangle}
\label{92}
\end{eqnarray}

 In the chosen reference frame and in the Bjorken limit 
($Q^2 \rightarrow \infty$, $x=Q^2/(2 P q)$) one obtains \cite{LPS1}
\begin{eqnarray}
&&W^{jl}=\delta^{jl}F_1(x,Q),\quad (j,l=1,2) \nonumber\\
&& W^{++}=\frac{1}{4(2-x)}[F_2(x,Q)-2xF_1(x,Q)].
\label{33'}
\end{eqnarray}

In the parton model the current operator is taken in IA, viz.
\begin{equation}
j^{\mu}(K\bm{e}_z;m,M')=
\langle 0,P^{+}|\Pi J_{free}^{\mu}(0) \Pi'|0,P^{'+}\rangle 
\label{90}
\end{equation}
where $\Pi$ and $\Pi'$ are the orthogonal projectors onto the
states with masses $m$ and $M'$, respectively (see Eqs. (\ref{free}, \ref{free'}). In Sect. \ref{S6}, we have already shown that the free current fulfills the extended Lorentz
covariance in our Breit frame; moreover, as already noted, the hermiticity property, Eq. (\ref{60}), does not impose any further constraint in DIS, since $m \not = M'$ \cite{LPS}. As far as the continuity equation is concerned, in the actual calculations of the structure functions only three components of the current are needed and can be chosen unconstrained with respect to the current conservation (e.g., one can use the $+$ and ${\perp}$ components of the free current operator in our special Breit frame, even in the case where the final state interaction is present), while the fourth component can be determined through the current conservation, see Eq. (\ref{67}) \cite{LPS1}.
 
  One of the major parton model assumptions is that the final state interaction (FSI) of the struck quark with the residual part of the target can be neglected. This implies that in our case the final states of the system, $|X \rangle$ are the states of two free particles:

\begin{equation}
|X\rangle =|p_1',\sigma_1'\rangle |p_2',\sigma_2'\rangle
\label{27}
\end{equation}

\noindent As a consequence within the parton model in the Bjorken limit one obtains $\xi=x$, and

\begin{eqnarray}
F_1(x) = \frac{1}{2}\rho(x)&=\frac{1}{2}&\sum_{\sigma_1,\sigma_2}\int\nolimits
|\chi_0({\bm k}_{\bot},\xi=x,\sigma_1,\sigma_2)|^2
\frac{d^2{\bm k}_{\bot}}{2(2\pi)^3x(1-x)}
\label{35'}
\end{eqnarray}

\noindent where  $\rho(x)$ is the probability distribution of the momentum fraction. Furthermore, all the longitudinal components of
$W^{\mu\nu}$ are equal to zero, i.e.,

\begin{equation}
W^{+\nu} = W^{-\nu} = W^{\mu+ }= W^{\mu-} = 0~~~,
\label{35''}
\end{equation}
and then the Callan-Gross relation holds

\begin{equation}
F_2(x) = 2xF_1(x)
\label{36'}
\end{equation}

 Let us now consider the exact hadronic tensor for the two particles,
which in our exactly solvable model interact via a relativistic harmonic
oscillator potential.

In the front-form dynamics if ${\tilde M}$ is the mass operator for the
functions $\Psi$ (see Eq. (\ref{5'})) and the interaction operator $V$  is defined as ${\tilde M}^2 = M_0^2+V$, then the equation ${\tilde M}^2\Psi_n = M_n^2\Psi_n$ has the same form as the nonrelativistic Schroedinger equation in momentum representation:

\begin{equation}
(\frac{{\bm k}^2}{m_0}+v)\Psi_n({\bm
k},\sigma_1,\sigma_2) = E_n\Psi_n({\bm k},\sigma_1,\sigma_2)
\label{13}
\end{equation}

\noindent where

\begin{equation}
v=V/4m_0,\quad E_n=(M_n^2-4m_0^2)/4m_0
\label{14}
\end{equation}

 We choose the function $v(r)$ in the form $v(r)=a^4r^2/m_0$, where $a$
is some constant with the dimension $GeV$. Then Eq. (\ref{13}) is the
well-known equation for the harmonic oscillator and the solutions are given by products of harmonic oscillator eigenstates and ordinary spin
eigenfunctions. The eigenvalues of the mass operator are equal to

\begin{equation}
M_n=2[m_0^2+a^2(2n+3)]^{1/2}
\label{16}
\end{equation}

\noindent with $n=n_1+n_2+n_3$ ( $n_i = 0,1,2...$ ).

  While in the parton model $F_1(x)$ and $F_2(x)$ are continuous functions of $x$, using the exact final state eigenfunctions the structure functions become linear combinations of delta-functions, which, at fixed $Q$, are not equal to zero only for discrete values of $x$. In order to recover continuous structure functions we consider average values over small intervals of $x$, which resemble the finite experimental resolution,

\begin{equation}
{\bar F}_i({\bar x},Q)=\frac{1}{x_2-x_1}\int_{x_1}^{x_2} F_i(x,Q)dx , \quad
i=1,2
\label{50}
\end{equation}

\noindent where ${\bar x}$ belongs to the small interval $[x_1,x_2]$,
such that $x_2-x_1\ll {\bar x}$.

\noindent We have demonstrated \cite{LPS1} that in the Bjorken limit
\begin{equation}
{\bar F}_1({\bar x},Q) \rightarrow {\bar F}_1({\bar x}) = \frac{1}{2} \rho({\bar
x}) , \quad  {\bar F}_2({\bar x},Q) \rightarrow {\bar F}_2({\bar x}) = 2 {\bar x}
{\bar F}_1({\bar x}). 
\label{54}
\end{equation}
Therefore, within our exactly solvable model, in the Bjorken limit the
averaged structure functions ${\bar F}_1$ and ${\bar F}_2$ do not depend on $Q$, namely one has Bjorken scaling, and the results of the parton model are recovered.
Furthermore the usual interpretation of the Bjorken variable $x$ as the momentum fraction of the struck quark ($\xi=x$) is also valid if the FSI is exactly taken into account.

\section{Deuteron electromagnetic form factors}
\label{S10}

For the evaluation of the elastic deuteron electromagnetic form factors we will adopt a current which satisfies extended Lorentz covariance, hermiticity and current conservation.

 Let $\Pi$ be the projector onto the subspace of bound states $|\chi \rangle$ of mass $m$ and spin $S$, and let ${\cal{J}}^{\mu}(K\bm{e}_z;m,m)$ be a current which fulfills extended Lorentz covariance. As we have already seen (see Eqs. (\ref{free}, \ref{free'}), a possible choice is the following one
\begin{eqnarray}
{\cal{J}}^{\mu}(K\bm{e}_z;m,m)=
\langle 0, P^+|\Pi J_{free}^{\mu}(0) \Pi| 0, P^{'+} \rangle 
\label{95}
\end{eqnarray}
where
\begin{equation}
P^{+}=\frac{1}{\sqrt{2}}[(m^2+K^2)^{1/2}+K],\quad
P^{'+}=\frac{1}{\sqrt{2}}[(m^2+K^2)^{1/2}-K]
\label{96}
\end{equation}
and $K=Q/2$. A choice for the current compatible with the hermiticity
condition, Eq. (\ref{60}), and with the extended Lorentz covariance is 
\cite{LPS}
\begin{eqnarray}
j^{\mu}(K\bm{e}_z;m,m)=\frac{1}{2}\{{\cal{J}}^{\mu}
(K\bm{e}_z;m,m) +
{\cal{J}}^{\mu}(-K\bm{e}_z;m,m)^* \}
\label{97}
\end{eqnarray}
where ${\cal{J}}^{\mu}(-K\bm{e}_z;m,m)$ is given by
\begin{eqnarray}
 && {\cal{J}}^{\mu}(-K\bm{e}_z;m,m) =
L^{\mu}_{\nu}[r_x(-\pi)]~\exp(\imath \pi S_x)\nonumber\\
&&{\cal{J}}^{\nu} (K\bm{e}_z;m,m)
\exp(-\imath \pi S_x)
\label{95'}
\end{eqnarray}

This current fulfills also the current conservation. Indeed, as mentioned in Section 3, in the elastic case the extended Lorentz covariance, together with hermiticity, imposes current conservation \cite{LPS}. 

It can be immediately obtained that

\begin{eqnarray}
&&\langle m S S_z| j^+(K\bm{e}_z;m,m) | m S S_z \rangle =
 \langle m S S_z| {\cal J}^+(K\bm{e}_z;m,m) | m S S_z \rangle,
 \label{106} \\
&& \langle m S S_z| j_x(K\bm{e}_z;m,m) | m S S'_z \rangle = {1 \over 2}
[ \langle m S S_z| {\cal J}_x(K\bm{e}_z;m,m)
| m S S'_z \rangle -  \nonumber\\
&&\langle m S S'_z| {\cal J}_x(K\bm{e}_z;m,m) | m S S_z \rangle ]
\end{eqnarray}

In the elastic case one has only $2S+1$ non-zero independent matrix elements for the em current defined in Eqs. (\ref{97}) and (\ref{95'}),
corresponding to the $2S+1$ elastic form factors. The independent matrix elements can be chosen as the diagonal matrix elements of $j^+$  with $S_z\geq 0$ and the matrix elements 
$\langle m S S_z| j_x(K\bm{e}_z;m,m) | m S S_z-1 \rangle$ of $j_x$ with 
$S_z\geq + 1/2$ \cite{LPS}.

For the deuteron the matrix elements of the current are related to the elastic form factors $F_0$, $F_1$, $F_2$ by the following general expression of the macroscopic current \cite{Glaser,riv}
\begin{eqnarray}
&&\langle m_d 1 S_z| j^{\mu}(K\bm{e}_z;m_d,m_d) | m_d 1 S_z' \rangle =
\nonumber \\ 
&&e^{*\alpha}_{1S_z}e^{\beta}_{1S'_z} \left\{ (P + P')^{\mu} 
\left[ - F_0 g_{\alpha \beta} - 
\frac{1}{2m_d^2} F_2 q_{\alpha} q_{\beta} \right] + 
F_1 \left( g_{\alpha}^{\mu}q_{\beta} - g_{\beta}^{\mu}q_{\alpha} \right)
\right\}
\label{100} 
\end{eqnarray}
where $e^{\alpha}_{1S_z}$ is the deuteron polarization vector and $q = P - P'$.

 A relevant result of our approach is that, if one adopts in the left hand side the microscopic current defined by the Eqs. (\ref{free'}, \ref{95}), (\ref{97}) and (\ref{95'}), the extraction of elastic em form factors is no more plagued by the ambiguities which are present when the free current is used in the reference frame where $q^+=0$ (see, e.g., \cite{GrKo,CCKP,KS}), as we discussed in the Introduction. For the deuteron only three of the matrix elements
$\langle m_d S S_z| j^{\mu}(K\bm{e}_z;m_d,m_d) | m_d S S_z' \rangle$
are independent (e.g., $\langle m_d 1 0| j^+ | m_d 1 0
\rangle$, $\langle m_d 1 1| j^+ | m_d 1 1 \rangle$, 
$\langle m_d 1 1| j_x | m_d 1 0 \rangle$), corresponding to the three elastic em form factors (if the current is taken free in the $q^+=0$ frame, one has four independent matrix elements \cite{GrKo}).

\begin{table}
\caption{Deuteron magnetic and quadrupole moments, corresponding to different $D$-state percentages, $P_D$ ( $\mu _d^{exp} = 0.8574$, $Q_d^{exp} = 0.2859 fm^2$ ).}
\begin{center}
\begin{tabular}{|c|c|c|c|c|c|}
\hline
Interaction & $P_D$  & $\mu _d$ (ref.\cite{CCKP}) & $\mu _d$ (this
paper)  & $Q_d$ (ref.\cite{CCKP}) & $Q_d$ (this paper) \\
\hline 
      &   &   &   &   &  \\
RSC       & 6.47   & 0.8500  & 0.8611  & 0.2804  & 0.2852 \\
Av14      & 6.08   & 0.8516  & 0.8608  & 0.2866  & 0.2907 \\
Paris     & 5.77   & 0.8531  & 0.8632  & 0.2795  & 0.2841 \\
Av18      & 5.76   &         & 0.8635  &         & 0.2744 \\
\hline
\end{tabular}
\end{center}
\label{table2}
\end{table}
We computed the form factors $A(Q^2)$ and $B(Q^2)$, which appear in the
unpolarized cross section, and the tensor polarization $T_{20}$. These quantities can be expressed as follows
\begin{eqnarray}
&&A(Q^2) = G_C^2+ \frac{8}{9} \eta ^2 G_Q^2+ \frac{2}{3} \eta G_M^2 \nonumber \\
&&B(Q^2) = \frac{4}{3} \eta ( 1 + \eta ) G_M^2 \nonumber \\
&&T_{20} = - \eta \frac{\sqrt{2}}{3} 
\frac{ \frac{4}{3} \eta G_Q^2 + 4 G_Q G_C + f G_M^2}{A + B \tan ^2(\theta / 2)}
\label{98'}
\end{eqnarray}
where $\eta = Q^2/(4 m^2_d)$, $f = 1/2 + (1 + \eta ) \tan ^2 (\theta / 2)$, and
\begin{eqnarray}
&&G_C(Q^2) =  (1 + \frac{2}{3} \eta ) F_0 - \frac{2}{3} \eta F_1
- \frac{2}{3} \eta ( 1 + \eta ) F_2 \nonumber \\
&&G_M(Q^2) = F_1 \nonumber \\
&&G_Q(Q^2) = F_0 - F_1 - ( 1 + \eta ) F_2
\label{98''}
\end{eqnarray}
with $G_C(0) = 1$, $G_Q(0) = m_d^2 Q_d$, and $G_M(0) = \mu _d m_d/m_p$.
The form factors $F_0$, $F_1$, and $F_2$ can be easily obtained from the matrix elements of the current in our Breit frame, since from Eq. (\ref{100}) one has
\begin{eqnarray}
&&\langle m_d 1 1| j^+(K\bm{e}_z;m_d,m_d) | m_d 1 1 \rangle = 
\sqrt{2} m_d (1 + \eta )^{\frac{1}{2}} F_0 \nonumber \\
&&\langle m_d 1 1| j_x(K\bm{e}_z;m_d,m_d) | m_d 1 0 \rangle = 
\sqrt{2} m_d (1 + \eta )^{\frac{1}{2}} \eta ^{\frac{1}{2}} F_1 
\nonumber \\
&&\langle m_d 1 0| j^+(K\bm{e}_z;m_d,m_d) | m_d 1 0 \rangle = \nonumber \\
&&= \sqrt{2} m_d (1 + \eta )^{\frac{1}{2}}
\left[ F_0 ( 1 + 2 \eta ) - 2 \eta F_1 - 2 \eta ( 1 + \eta ) F_2 \right]
\label{98'''}
\end{eqnarray}

 Let us first study the form factors at $Q^2=0$, namely the deuteron magnetic  moment, which in nuclear magnetons is given by
\begin{equation}
\mu _d = \lim_{Q \rightarrow 0}  \sqrt{2} m_p 
\langle m_d 1 1| j_x(K\bm{e}_z;m_d,m_d) | m_d 1 0 \rangle 
 /(Q m_d)
\label{99'}
\end{equation}
and the deuteron quadrupole moment
\begin{eqnarray}
 Q_d = \frac{\sqrt{2}}{(Q^2 m_d)}  
\lim_{Q \rightarrow 0}   
[& \langle m_d 1 0| j^+(K\bm{e}_z;m_d,m_d) | m_d 1 0 \rangle \nonumber \\
-& \langle m_d 1 1| j^+(K\bm{e}_z;m_d,m_d) | m_d 1 1 \rangle ],
\label{101}
\end{eqnarray}
which require only the knowledge of the deuteron wave function and do not involve nucleon form factors. Our results corresponding to different $N-N$ interactions are compared in Table \ref{table2} with the results of Ref. \cite{CCKP}, obtained using the free current in the $q^+=0$ reference frame.

 We obtain an increase of the order of 2\% with respect to the non
relativistic results, both for $\mu _d$ and $Q_d$, while the results of
Ref. \cite{CCKP} show an increase of the order of 1\% for $\mu _d$ and are essentially equal to the non relativistic ones for $Q_d$. As a consequence, for  $\mu _d$ our covariant approach prefers high $P_D$ values, while the $q^+=0$ approach points to a low $P_D$. For $Q_d$ our results corresponding to the RSC interaction, which has the highest $P_D$ value, is the closest to the experimental value.

In Figs. 1, 2, and 3 we report our results for $A(Q^2)$, $B(Q^2)$, and $T_{20}(Q^2)$, respectively.
These quantities have been calculated with different realistic $N-N$ interactions (Reid soft core \cite{RSC}, AV14 \cite{AV14}, AV18 \cite{AV18}, and Paris \cite{Paris}), using the H\"{o}hler \cite{Hoehler} nucleon form factors in Figs. 1(a), 2(a), 3(a) and the Gari-Kr\"{u}mpelmann \cite{Gari} nucleon form factors in Figs. 1(b), 2(b), 3(b). 

\begin{figure}
\epsfig{bbllx=12mm,bblly=206mm,bburx=0mm,bbury=284mm,file=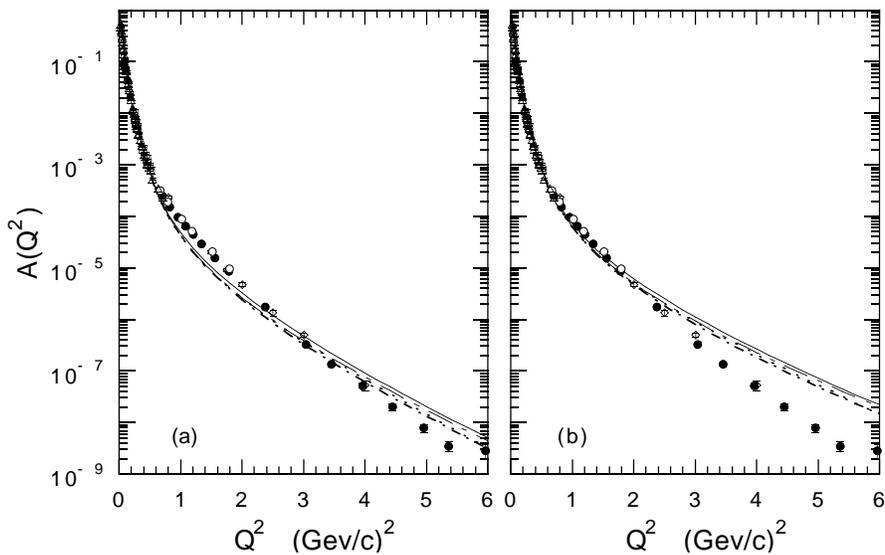}
\caption[ ]{(a) The deuteron form factor $A(Q^2)$ obtained with the nucleon form factors of Ref. \cite{Hoehler} and different $N-N$ interactions: RSC (solid line), AV14 (dot-dashed line), AV18 (long dashed line), Paris (dotted line). Experimental data are from Ref. \cite{Galster} (squares), Ref. \cite{SLAC} (diamonds), Ref. \cite{Saclay} (triangles), Ref. \cite{JLABA} (full dots) and \cite{JLABC} (open dots). (b) The same as in (a), but for the nucleon form factors of Ref. \cite{Gari}.}
\label{A}
\end{figure}

The overall behaviour of the available experimental data is reproduced, but the dependence of the results on the nucleon form factors is remarkable for $A(Q^2)$ and $B(Q^2)$, especially at high $Q^2$. For $A(Q^2)$, as shown in Fig. 1, this dependence is much stronger than the effect of different $N - N$ interactions.  
Because of this well known dependence on different parametrizations of the nucleon form factors, which reflects our ignorance of the nucleon electromagnetic structure, the deuteron form factors, calculated within non-covariant theories,  have been used to gain information on the nucleon form factors (see, e.g., \cite{Galster,Saclay}).
Our calculation could be used as a benchmark to this end, since it is based on a current which satisfies all the requirements of extended Poincar\'e covariance, hermiticity and current conservation. We think that the uncertainties about the nucleon form factors should be cleared up, before introducing explicitly two-body currents with the aim to obtain a more precise description of the data. In any case the two-body currents will have to fulfill separately the constraints of extended Poincar\'e covariance and hermiticity, as the current defined in Eqs. (\ref{95}, \ref{97}) does. 

The tensor polarization shows an higher dependence on the interaction and, once again, an high $P_D$ value looks better. An investigation of the effect of possible changes in the deuteron wave function should be performed.
A more stringent comparison with the new TJNAF data for $B(Q^2)$ and $T_{20}(Q^2)$ will be possible in the near future.

\begin{figure}
\epsfig{bbllx=12mm,bblly=206mm,bburx=0mm,bbury=284mm,file=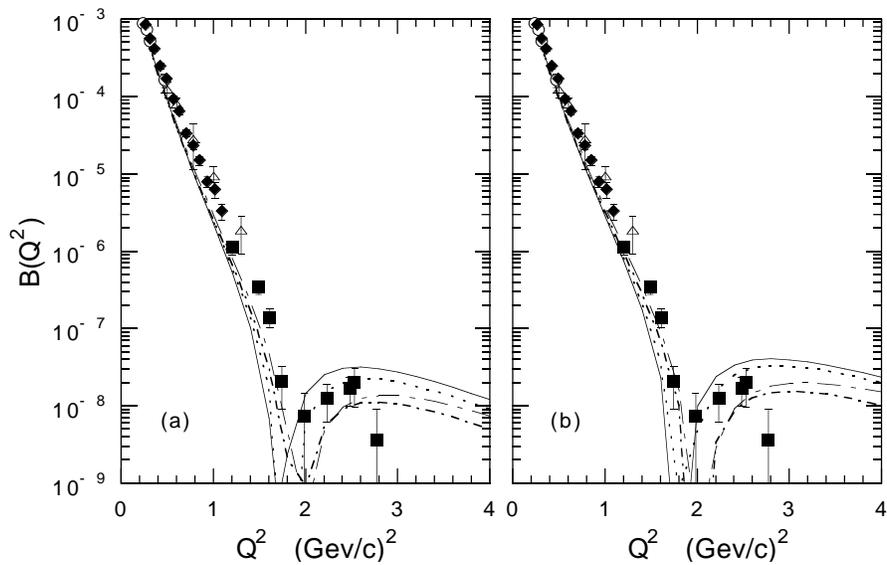}
\caption[ ]{(a) The deuteron form factor $B(Q^2)$ obtained with the nucleon form factors of Ref. \cite{Hoehler} and different $N-N$ interactions, as in Fig. 1 (a). Experimental data are from Ref. \cite{Buchanan} (open dots), Ref. \cite{Ganichot} (open squares), Ref. \cite{Sac-Auffret} (diamonds), Ref. \cite{Cramer} (triangles) and Ref. \cite{SLAC-Bosted} (full squares).(b) The same as in (a), but for the nucleon form factors of Ref. \cite{Gari}.}
\label{B}
\end{figure}
\begin{figure}
\epsfig{bbllx=12mm,bblly=206mm,bburx=0mm,bbury=284mm,file=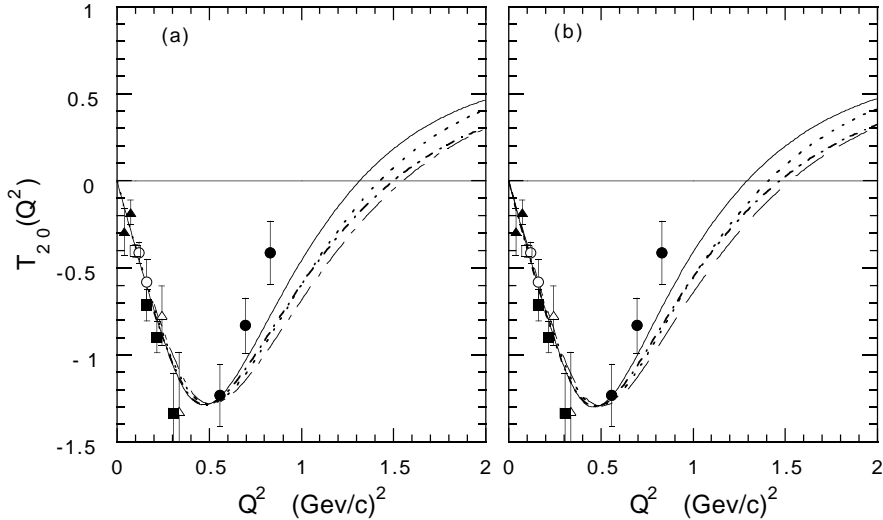}
\caption[ ]{(a) The tensor polarization $T_{20}(Q^2)$ obtained with the nucleon form factors of Ref. \cite{Hoehler} and different $N-N$ interactions, as in Fig. 1 (a). Experimental data are from  Ref. \cite{Bates84} (open dots), Ref. \cite{Novo85} (full triangles), Ref. \cite{Novo90} (open triangles), Ref. \cite{The} (full dots), Ref. \cite{Ferro} (open squares), and Ref. \cite{NIKHEF} (full squares). }
\label{T}
\end{figure}

\section{ Conclusion}
\label{S11}

 The constraints imposed by extended Poincar\'e covariance and current conservation allow one to determine the current operator through some auxiliary operators $j^{\mu}(K{\bm e}_z;M_i,M_j)$. These auxiliary operators are obtained by  projecting the current into subspaces of definite mass and spin values, and then evaluating  matrix elements between total momentum eigenstates in the Breit frame where the momentum transfer is directed along the spin quantization axis. The auxiliary operators act only through internal variables and have to be covariant for rotations around the $z$ axis. We have shown that it is possible to choose explicit models for $j^{\mu}(K{\bm e}_z;M_i,M_j)$, such that all the necessary requirements are satisfied. In particular, as noted in
Sects. \ref{S6}, \ref{S9} and \ref{S10} (see especially Eqs. (\ref{90}), (\ref{95}) and (\ref{97})), the operator $j^{\mu}(K{\bm e}_z;M_i,M_j)$ can be obtained by projecting the free current operator onto the subspaces corresponding to definite eigenvalues of the mass and spin operators. It is worth mentioning that the second term in Eq. (\ref{97}), introduced to ensure the current hermiticity for elastic scattering, is implicitly an interaction dependent term, since the rotations around the $x$ axis are interaction dependent in the front form.

  In Sect. \ref{S9} we have considered the application of our results to
DIS in an exactly solvable model. We consider a model system of two relativistic particles interacting via the relativistic harmonic oscillator potential and adopt an electromagnetic current operator, whose matrix elements exhibit the correct properties under Poincar\'e transformations. Then, in the framework of the front-form hamiltonian dynamics, we can derive exact expressions for the DIS structure functions, including the FSI effects, and show that in the Bjorken limit the exact results coincide with those given by the parton model, after an average over small intervals of the scaling variable $x$ has been performed (this average features the finite detector resolution).

  In Sect. \ref{S10} we have applied our results to the calculation of the deuteron elastic form factors. In contrast with the approaches discussed in the Introduction, we have no problem with the angular condition. Indeed our model current is in agreement with extended Poincar\'e covariance and current conservation, by construction, and therefore the number of independent matrix elements of the current is equal to the number of physical form factors.

Our approach, based on the reduction of the whole complexity of the Poincar\'e covariance to the SU(2) symmetry can represent a simple framework where to investigate the possible many-body terms to be added to the free current, since they must obviously fulfill the rotational covariance condition of Eq. (\ref{56}).


\end{document}